\documentclass[a4paper,11pt]{article}
\usepackage{pos}
\usepackage{amsmath}

\title{Multi-Messenger Search for Neutrino and Gravitational-Wave Emissions from Binary Black Holes Near Active Galactic Nuclei}
\ShortTitle{Neutrino and Gravitational-Wave Emissions from Binary Black Holes Near AGNs}

\author*[a]{Leonardo Ricca}
\author[a]{Matthias Vereecken}
\author[a]{Christoph Raab}
\author[a]{Mathieu Lamoureux}
\author[a]{Giacomo Bruno}
\author[a]{Gwenhaël De Wasseige}

\affiliation[a]{Centre for Cosmology, Particle Physics and Phenomenology - CP3, Université Catholique de Louvain,
\\
Chemin du Cyclotron 2, Louvain-La-Neuve, Belgium}

\emailAdd{leonardo.ricca@uclouvain.be}

\abstract{Binary black holes (BBHs) in the vicinity of Active Galactic Nuclei (AGNs) are particularly interesting systems from both a cosmological and astrophysical point of view. Matter and radiation fields within the dense AGN environment could produce electromagnetic and neutrino emission in addition to gravitational waves (GWs). Moreover, interactions between BBHs and AGN accretion disks are expected to influence BBH formation channels and merger rates. Understanding these sources could help explain the unexpectedly high BBH masses observed through GWs by the LIGO-Virgo-KAGRA collaborations. We present a search for coincident gravitational-wave and neutrino emission from AGNs. Our innovative approach combines information from gravitational-wave data, neutrino observations, and AGN optical catalogs to increase the chances of identifying potential sources and studying their properties. We assess the sensitivity of the search using subthreshold gravitational-wave candidates from LIGO-Virgo-KAGRA data, neutrino event candidates from public IceCube Neutrino Observatory data and AGN candidates from the Quaia catalog. A confident detection of such an event would mark a breakthrough in multi-messenger astronomy.}

\ConferenceLogo{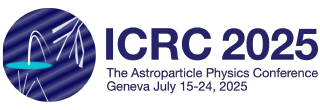}

\FullConference{39th International Cosmic Ray Conference (ICRC2025)\\
 15–24 July 2025\\
Geneva, Switzerland\\}

\begin{document}
\maketitle

\section{Introduction}
On August 17, 2017, the detection of a gravitational-wave signal from a binary neutron star merger coincident with a gamma-ray burst marked the first multi-messenger observation involving gravitational waves. No associated neutrino signal was detected for this event.
Numerous scenarios have been proposed in which a joint gravitational wave (GW) and neutrino emission is produced.
This has motivated several follow-up searches for neutrino counterparts to gravitational-wave detections, using independent methods and covering different neutrino energy ranges. However, no evidence of an association between gravitational waves and neutrinos has been found so far \cite{IceCubeO3, GRECOO3, ELOWENO4a}. \\ 
Multi-messenger events of this kind are generally expected from mergers involving at least one neutron star, as the presence of accreting matter in these systems enables particle acceleration, interaction, and potentially electromagnetic or neutrino emission.\\
Nevertheless, theoretical models have been proposed suggesting the potential for multi-messenger emission from binary black holes (BBHs) mergers, motivated by tentative detections of electromagnetic counterparts \cite{Perna16, Graham20}.
One class of models considers BBH mergers occurring within the accretion disks of active galactic nuclei (AGNs) \cite{McKernan:2019hqs, Kimura:2021xxu, Tagawa:2023uqa}. These environments are expected to host a rich population of black holes, providing favorable conditions for binaries to form and merge. Additionally, this scenario represents a promising formation channel for the BBH mergers observed by the LIGO, Virgo, KAGRA (LVK) collaborations, as it allows for hierarchical mergers. These are consecutive black hole coalescences, that could account for the unexpectedly high black hole masses reported in gravitational-wave detections \cite{LVK23, Yang_2019}.
From a multi-messenger perspective, what makes these sources promising is the possibility that matter from the AGN accretion disk may accrete onto the remnant of the BBH merger, potentially leading to the emission of electromagnetic and/or neutrino signals.\\
We present here a method to search for coincident neutrino and GW signals emitted by BBHs within AGNs accretion disks. We will study the sensitivity of the search using GW candidates from LIGO-Virgo-KAGRA GWTC3 catalog \cite{GWTC3}, neutrino event candidates from publicly available IceCube Neutrino Observatory data for point-source searches \cite{IceCube:2021} and AGNs candidates from the Quaia catalog \cite{Quaia}.
The requirement of spatial coincidence with the position of a known AGN, both in sky direction and distance, is the key advantage of the proposed method, since it effectively reduces the rate of spurious associations between noise fluctuations in gravitational wave and neutrino data. As a result, the analysis can incorporate lower-significance GW signal triggers, i.e. candidates with higher false alarm rates (FAR). Indeed, even if some of the individual messengers observations are less significant, detecting them in coincidence could sufficiently enhance the significance of their combined observation.\\
The inclusion of subthreshold GW triggers, combined with a new statistical framework designed to analyze these events mitigating contamination from noise-like triggers, represent the main advancement respect to \cite{Bruno23}, where a search for potential neutrino and GW associations coming from the same kind of sources was presented. 


\subsection{Data}
The inputs to our analysis consist of candidate events from each messenger—gravitational waves and neutrinos—as well as a catalog of active galactic nuclei candidates.\\
For neutrino candidates we consider the public neutrino dataset provided by the IceCube collaboration in the all-sky point-source 10 years data release (from 2008 to 2018) \cite{IceCube:2021}. The catalog is composed of track-like events, which are characterized by their excellent direction reconstruction capabilities and small angular uncertainties.
It is all-sky, with an event rate of roughly 4 mHz and is dominated by background, consisting mainly of atmospheric muon neutrinos in the Northern sky and atmospheric muons in the Southern sky.\\
The third LIGO Scientific, Virgo, and KAGRA (LVK) Collaboration Gravitational-Wave Transient Catalog (GWTC-3) \cite{GWTC3} represents the most comprehensive dataset of GW observations publicly available so far. In addition to significant candidates, which are defined by an inferred probability of astrophysical compact binary coalescence (CBC) origin of $p_{\text{astro}} > 0.5$, we also include in the analysis the sample of subthreshold events listed in the catalog. These triggers do not meet the criteria required for $p_{\text{astro}}$, however they are selected because of their fairly low false alarm rate (FAR) $<2.0 \; \text{day}^{-1}$. In this work, we consider only signal candidates detected during the third LIGO-Virgo observing run (O3), conducted between 2019 and 2020. Although this data sample does not temporally overlap with the IceCube point-source catalog, this does not affect the aim of our study, as we only want to perform a sensitivity study rather than a real search. For our purposes, it is therefore sufficient to time-shift the neutrino dataset to cover the LVK O3 observing run.\\\
Our set of AGN candidates is derived from the Quaia catalog \cite{Quaia}. Quaia is the AGN catalog that samples the largest comoving volume among all existing spectroscopic quasar datasets and is characterized by its good all-sky coverage and high completeness. Candidates are obtained by crossmatching the broader, though less pure, Gaia DR3 quasar catalog with unWISE infrared data.
The catalog contains 185,831 AGNs with magnitude $G<19$, of which 109,005 have an estimated redshift smaller than 1.5, making them significant for searching gravitational wave associations.

\section{Analysis method}
We propose a method to assess the significance of a multi-messenger association involving a gravitational wave trigger, a sample of neutrino candidates and an AGN from a selected catalog. This is performed by computing a ranking statistic for the proposed association and comparing it with its distribution for the null hypothesis, thus obtaining a p-value.\\
Initially, the triggers are generated independently from the gravitational-wave data and the neutrino data. 
Given that in this search we allow ourselves to consider also GW candidates with low statistical significance, it is crucial that the ranking statistic incorporates our knowledge of the individual confidence levels of each messenger. Hence we consider the following hypotheses: ($H^C$) both datasets contain transient signals originating from a common source, spatially consistent with the location of a known AGN; ($H^{NN}$) both datasets contain only noise; ($H^{SN}$), ($H^{NS}$) one of the two datasets (respectively LVK or IceCube) contain a transient signal, while the other contains only noise; and ($H^{SS}$) both datasets contain signals, but these signals originate from independent sources. 

The joint ranking statistic used in this analysis is defined as the Bayes factor contrasting the hypothesis of interest $H^C$ to the logical disjunction of all alternative hypotheses:
\begin{equation}
    \Lambda = \frac{P(D_{\text{LVK}},D_{\text{IC}}|H^C)}{P(D_{\text{LVK}},D_{\text{IC}}|H^{NN}\lor H^{SN}\lor H^{NS})}
\end{equation}

where $D_{\text{LVK}}$ and $D_{\text{IC}}$ are the datasets from LIGO-Virgo-KAGRA and IceCube.

After performing mathematical manipulations analogous to those outlined in \cite{Ashton_2018, Stachie_2020, Pillas_2023} we can rewrite the ranking statistic as:

\begin{equation}
    \Lambda = \frac{I_{\vec{r}}}{1 + Q_{\text{LVK}} + Q_{\text{IC}} + Q_{\text{LVK}} Q_{\text{IC}}}
\end{equation}

where $Q_{D}$, with $D=\text{LVK}, \text{IC}$, are the inverse Bayes factors of the individual messengers, comparing the noise-only hypothesis ($H^N$) to the signal-plus-noise hypothesis ($H^S$) in the LVK and IceCube datasets, respectively:

\begin{equation}
    Q_{D} = \frac{P(D|H^N)}{P(D|H^S)}
\end{equation}

At the numerator, $I_{\vec{r}}$ is an integral that measure the overlap of the posterior distributions of the individual messengers over the common parameters; in this study, we focus only on the inferred source position. If we call $r$ the distance of the source, $\Omega$ the sky location, we can write more explicitly:

\begin{equation}
        I_{\vec{r}} = \iint_{\{\vec{r},P(\vec{r}|H^C)>0\}} dr d\Omega \frac{P(r,\Omega|D_{\text{LVK}},H^S)P(r,\Omega|D_{\text{IC}},H^S)}{P(r,\Omega|H^S)}
\end{equation}

As discussed in \cite{Ashton_2018}, the domain of this integral is restricted to the prior support of $H^C$. Given that in our analysis the common source hypothesis is conditioned on the AGN emission model hypothesis, the prior support can be reformulated as the set of positions consistent with a known AGN, i.e. $P(r,\Omega|H^{\text{AGN}})>0$. In our work, the probability $P(r,\Omega|H^{\text{AGN}})$ is given by the Quaia catalog. Given that AGNs have far more precise sky localizations than the GW and neutrino triggers, we approximate their positions as delta functions and consequently rewrite the integral as:

\begin{equation}
    I_{\vec{r}} = N \sum_{i=1}^NP(\vec{r}_{\text{AGN}_i}|D_{\text{LVK}},H^S)P(\vec{r}_{\text{AGN}_i}|D_{\text{IC}},H^S)
\end{equation}

where $N$ is the total number of AGNs in the catalog. Here we assume that each AGN has an equal prior probability of hosting a source and this gives the factor $N$ multiplying the sum.
To search for potential associations, we select neutrino candidates according to a time window around the time of the GW trigger.
Theoretical models \cite{McKernan:2019hqs, Kimura:2021xxu, Tagawa:2023uqa} suggest that, for BBH mergers occurring within AGN accretion disks, a time delay between the emission of gravitational waves and neutrinos is expected. However, the magnitude of this offset varies depending on the physical scenario considered. Moreover, these estimates depend on source parameters that are typically unknown or difficult to infer from GW observations alone. These studies indicate that neutrino emission is expected to occur after the merger, with a time offset $\mathcal{O}(10 \: \text{days})$. 
For this reason we choose a long and asymmetric time-window of $[-1,100]$ days around the GW trigger merger time, which encompass a wide range of potential time delays



The inverse Bayes factor $Q_{\text{LVK}}$ and the spatial posterior distribution $P(r,\Omega|D_{\text{LVK}},H^S)$ for GW data are computed using BAYESTAR \cite{bayestar}. BAYESTAR is a Bayesian localization algorithm which can generate, within seconds, a reliable skymap, without sampling the intrinsic source parameters but achieving an accuracy comparable to that of full parameter estimation.
For each GW trigger, BAYESTAR computes two Bayes factors: one comparing the signal-plus-noise hypothesis against the noise-only hypothesis, and another evaluating the coherent GW signal against the incoherent noise transients hypothesis. We use the former as $Q_{\text{LVK}}$. The spatial posterior distribution is computed on an adaptive HEALPix grid, which has higher resolution in regions of the sky with higher posterior probability.

Regarding IceCube data, we have developed a method to perform a fast Bayesian inference and compute the evidence $P(D_{\text{IC}}|H^S)$ and the posterior distribution $P(r,\Omega|D_{\text{IC}},H^S)$. A fast method is required because performing Bayesian inference over the full sky and a 100-day time window can be too computationally intensive with standard techniques. The likelihood employed is the neutrino point-source likelihood:

\begin{equation}
\label{eq:nu_likelihood}
\mathcal{L}_{S+B}(\vec{x}; n_s) = \prod_{\nu_i} \left[ \frac{n_s}{N} \mathcal{S}(\vec{x}; \vec{x}_{\nu_i}) + \left( 1 - \frac{n_s}{N} \right) \mathcal{B}(\vec{x}) \right]
\end{equation}

where $\nu_i$ are the neutrinos observed within the time window under consideration, $N$ is the total number of observed neutrinos and $n_S$ the inferred number of signal neutrinos. $\mathcal{S}$ and $\mathcal{B}$ represent the signal and background probability density functions, respectively. For this analysis, we consider only their spatial dependence. The signal term represents the probability of observing a neutrino from a direction $\vec{x}_{\nu_i}$, given a true source direction $\vec{x}$, and is modeled using the von Mises–Fisher distribution, which is an extension of the Gaussian distribution to spherical geometry. The background distribution is inferred directly from the neutrino data by fitting over the entire available dataset, and depends only on declination.\\

Since computing the test statistic requires the posterior distribution over distance, we formulate the priors in terms of the source distance $r$ and neutrino luminosity $L$. We adopt physically motivated priors: uniform in volume (i.e., uniform over sky direction and proportional to $r^2$ in distance), and log-uniform over several orders of magnitude in source luminosity.

\begin{align}
    \Pi(\Omega) \propto \text{const}, \quad &\text{for} \quad \Omega \in S^2\\
    \Pi(r) \propto r^2, \quad &\text{for} \quad r \in [10,6000] \; \text{Mpc}\\
    \Pi(\log L) \propto \text{const}, \quad &\text{for} \quad \log L \in [48,54] \; \text{erg}
\end{align}

The choice of the upper bound of the $\log L$ prior is motivated by current upper limits on the isotropic equivalent energy emitted in neutrinos from GW events, which lie in the range $\log L \simeq 52 - 56 \; \text{erg}$ \cite{IceCubeO3, GRECOO3}. The $r^2$ dependence of the prior naturally favors more distant and brighter sources; however, sources with such high luminosity are likely less common in realistic astrophysical scenarios. It may therefore be beneficial to reduce their weight in the inference, for example, by choosing a lower upper bound for the $\log L$ prior or by implementing a non-uniform prior that decays for higher luminosities. We adopt the current uniform $\log L$ prior for simplicity, leaving a more detailed study of this issue for a future work.

The source distance $r$ and luminosity $L$ can be used to estimate the expected number of observed signal neutrinos $n_S$ given the detector's effective area and assuming a model for the astrophysical neutrino flux. 
\begin{equation}
    P(\hat{r}, d_L, \log L | D_{\nu}) \propto \mathcal{L}_{S+B} \left( \hat{r}, n_s = C \cdot \int A_{\text{eff}}(E_{\nu}, \delta) \, dE_{\nu} \cdot \frac{L}{4\pi d_L^2} \right) \times \Pi(d_L) \Pi(\log L)
\end{equation}

where $C$ is a normalization constant for the neutrino spectrum. In this study we consider a neutrino flux described by a power law $E^{-\gamma}$ over the energy range $10^2-10^8$ GeV, with spectral index $\gamma=2$. 

To speed up the computation we factorize the Bayesian inference by separating the posterior dependence on the sky location from that on the signal amplitude (or alternatively, distance and luminosity of the source). We hence divide the sky in a grid and evaluate the posterior on distance and luminosity for each pixel.
 In addition, we adopt an adaptive HEALPix grid, in which only regions of the sky with higher posterior probability are sampled at finer resolution. The inference begins with a scan over a coarse grid, followed by a refinement in the most relevant pixels.


\section{Search sensitivity}

\begin{figure}[h!]
    \centering
    \includegraphics[width=0.8\linewidth]{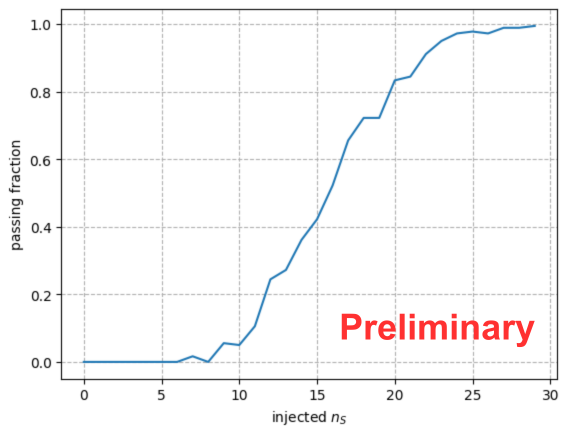}
    \caption{Fraction of events with ranking statistics above the 99th-percentile background threshold as a function of the number of injected neutrinos $n_S$}
    \label{fig:sensitivity}
\end{figure}

We assess the sensitivity of the search by injecting simulated neutrinos in temporal and spatial coincidence with the gravitational-wave event GW190521. This event has drawn significant interest because of the exceptionally high masses of the merging black holes and its tentative association with an optical counterpart, both of which may suggest a possible origin within an AGN accretion disk \cite{GW190521_astrophysical, Graham20}.
For the injection’s sky coordinates, we choose the position of AGN J124942.3+344929, which hosts the source of the proposed optical counterpart ZTF19abanrhr. As previously stated, the neutrino dataset we are considering does not cover the time of this GW detection. Since our goal is only to perform a sensitivity study rather than an actual search, it is sufficient to shift the neutrino dataset in time to have the necessary overlap.

We estimate the null-hypothesis distribution of the ranking statistic, corresponding to the case where no true association among gravitational-wave events, neutrinos, and AGN positions is present in the data, by time-scrambling the GW and neutrino datasets and randomizing the AGN coordinates. The random AGN catalog is produced so as to match the selection function of the original sample, as explained in \cite{Quaia}.

Finally, we calculate the fraction of injected events whose ranking-statistic values exceed the 99th percentile of the background distribution, as a function of the number of injected neutrinos $n_S$.
The sensitivity curve rises monotonically with $n_S$ and the observed fluctuations would smooth out with a larger sample of injected events.

The present study serves as a proof of concept for our proposed method to search for common sources of gravitational waves and neutrinos within an AGN environment. Observing a confident association between signals in these different messengers would constitute an important milestone for multi-messenger astronomy. 
In case no association is detected, and the sensitivity curve is shown to rise monotonically with $n_S$, our method can be used to place an upper limit on the neutrino luminosity of this class of events.


\section*{Acknowledgements}

This research has made use of data or software obtained from the Gravitational Wave Open Science Center (gwosc.org), a service of the LIGO Scientific Collaboration, the Virgo Collaboration, and KAGRA. This material is based upon work supported by NSF's LIGO Laboratory which is a major facility fully funded by the National Science Foundation, as well as the Science and Technology Facilities Council (STFC) of the United Kingdom, the Max-Planck-Society (MPS), and the State of Niedersachsen/Germany for support of the construction of Advanced LIGO and construction and operation of the GEO600 detector. Additional support for Advanced LIGO was provided by the Australian Research Council. Virgo is funded, through the European Gravitational Observatory (EGO), by the French Centre National de Recherche Scientifique (CNRS), the Italian Istituto Nazionale di Fisica Nucleare (INFN) and the Dutch Nikhef, with contributions by institutions from Belgium, Germany, Greece, Hungary, Ireland, Japan, Monaco, Poland, Portugal, Spain. KAGRA is supported by Ministry of Education, Culture, Sports, Science and Technology (MEXT), Japan Society for the Promotion of Science (JSPS) in Japan; National Research Foundation (NRF) and Ministry of Science and ICT (MSIT) in Korea; Academia Sinica (AS) and National Science and Technology Council (NSTC) in Taiwan.

\bibliographystyle{ICRC}
\bibliography{references}

%
%
%

\end{document}